\documentclass[conference]{IEEEtran}

\usepackage{url}
\usepackage{graphicx}
\usepackage{xcolor}
\usepackage{todonotes}
\usepackage{balance}
\usepackage[inline]{enumitem}

\usepackage[noabbrev,capitalize]{cleveref}

\usepackage[noadjust]{cite}

\def\paragraph#1{\subsubsection*{#1}}

\begin{document}
% todo: remove after shortening
% \linenumbers

% \title{OMG: On Security and Safety of Edge-AI in Large-Scale Scenarios}
%\title{Oh my Edge! On Security and Safety of Edge-AI in Large-Scale Scenarios}
\title{SoK: Towards Security and Safety of Edge AI}

\author{\IEEEauthorblockN{Anonymous Authors}}

\author{%
\IEEEauthorblockN{Tatjana Wingarz\IEEEauthorrefmark{1}, Anne Lauscher\IEEEauthorrefmark{1}, Janick Edinger\IEEEauthorrefmark{1}, Dominik Kaaser\IEEEauthorrefmark{2}, Stefan Schulte\IEEEauthorrefmark{2}, Mathias Fischer\IEEEauthorrefmark{1}}%
 %\and%
 %\IEEEauthorblockN{Janick Edinger\IEEEauthorrefmark{1}}%
 %\and%
 %\IEEEauthorblockN{Mathias Fischer\IEEEauthorrefmark{1}}%
 %\and%
 %\IEEEauthorblockN{Dominik Kaaser\IEEEauthorrefmark{2}}%
 %\and%
 %\IEEEauthorblockN{Anne Lauscher\IEEEauthorrefmark{1}}%
 %\and%
 %\IEEEauthorblockN{Stefan Schulte\IEEEauthorrefmark{2}}%
 %\linebreakand%
 \IEEEauthorblockA{\IEEEauthorrefmark{1} \textit{University of Hamburg}, Germany, \emph{\{firstname.lastname\}}@uni-hamburg.de \\
 \IEEEauthorblockA{\IEEEauthorrefmark{2} \textit{TU Hamburg}, Germany, 
 \emph{\{firstname.lastname\}}@tuhh.de}
 }
 }

% \author{Tatjana Wingarz}
% \affiliation{%
%   \institution{Universität Hamburg}
%   \country{Germany}}
% \email{tatjana.wingarz@uni-hamburg.de}

% \author{Janick Edinger}
% \affiliation{%
%   \institution{Universität Hamburg}
%   \country{Germany}}
% \email{janick.edinger@uni-hamburg.de}

% \author{Mathias Fischer}
% \affiliation{%
%   \institution{Universität Hamburg}
%   \country{Germany}}
% \email{mathias.fischer@uni-hamburg.de}

% \author{Dominik Kaaser}
% \affiliation{%
%   \institution{TU Hamburg}
%   \country{Germany}}
% \email{dominik.kaaser@tuhh.de}

% \author{Anne Lauscher}
% \affiliation{%
%   \institution{Universität Hamburg}
%   \country{Germany}}
% \email{anne.lauscher@uni-hamburg.de}

% \author{Stefan Schulte}
% \affiliation{%
%   \institution{TU Hamburg}
%   \country{Germany}}
% \email{stefan.schulte@tuhh.de}

\maketitle
\thispagestyle{plain}
\pagestyle{plain}

\begin{abstract}
Advanced AI applications have become increasingly available to a broad audience, e.g., as centrally managed large language models (LLMs).
Such centralization is both a risk and a performance bottleneck -- Edge AI promises to be a solution to these problems.
However, its decentralized approach raises additional challenges regarding security and safety.
In this paper, we argue that both of these aspects are critical for Edge AI, and even more so, their integration.
Concretely, we survey security and safety threats, summarize existing countermeasures, and collect open challenges as a call for more research in this area.
\end{abstract}

\begin{IEEEkeywords}
Edge AI, Security, Privacy, Safety
\end{IEEEkeywords}

%\todo[]{move footnotes to references or smth similar}

\section{Introduction}
Artificial Intelligence (AI) and machine learning (ML) are gaining huge interest from industry and society, with applications deployed in various areas, from autonomous driving to omnipresent speech recognition. 
% - Problem statement
Despite their impact, the criticism towards AI and ML is multifaceted \cite{huang2022overview}. 

% -- Societal dimension
\textit{From a societal perspective}, AI introduces the tendency to form monopolies as it requires large amounts of data. ML expertise thus accumulates at big companies like OpenAI, Google, or Meta as they have enough resources to collect data and train large AI models. Using these technologies typically requires sharing data with them, resulting in data privacy concerns and users losing data sovereignty over potentially sensitive information. Furthermore, current AI suffers from poor explainability and bias in training data, requiring additional safeguards that are challenging to implement \cite{lauscher2020general}. With the big companies as gatekeepers, such measures might even lead to censorship.

% -- Technical dimension
% --- bandwidth and latency
\textit{From a technical perspective}, uploading data to a centralized entity is not always possible due to bandwidth limitations and due to violations of timing constraints when (near-)real-time inference is required. 
% --- Performance bottleneck, SPoF
Further, centralized AI constitutes a performance bottleneck and a single point of failure.

% -- Edge AI our saviour
Moving AI to the network's edge can help to mitigate these problems. Edge AI refers to deploying AI algorithms and models directly on edge devices like smartphones and IoT devices. By performing computations locally, Edge AI reduces latency, preserves bandwidth, and enhances privacy. This approach is beneficial for applications requiring real-time decision-making or operating in environments with limited or intermittent connectivity to the Internet \cite{bornholdt2023measuring}. At the same time, Edge AI introduces new challenges: due to its distributed nature, control over AI-based algorithms diminishes, and the potential for attacks increases. The decentralization implies that AI models are deployed across many devices, each potentially vulnerable. As a result, security measures must be robustly implemented at each edge node to mitigate the risk of unauthorized access, tampering, or malicious exploitation, requiring inexpensive and scalable safeguards against various attacks.
% -- Resume of problem statement

% - Other surveys
To the best of our knowledge, existing surveys cover topics of general challenges of Edge AI \cite{shi2020communication,ding2022roadmap,iftikhar2023ai,singh2023edge}, focus on general AI security \cite{li2018cyber,mothukuri2021survey,oseni2021security} or safety \cite{hendrycks2021unsolved,mohseni2022taxonomy} but do not consider the intersection of those areas in the context of Edge AI. 
There are only two exceptions. First, the authors of \cite{sachdev2020towards} cover security/privacy aspects in the context of Edge AI, but are focused on the subdomain of digital marketing environments and do not consider a broader application. Second, the authors of \cite{ansari2020security} outline some security threats to Edge AI, but their work is limited in scope and does not cover any safety implications. 
Finally, the safety definition used by existing surveys on AI safety \cite{hendrycks2021unsolved,mohseni2022taxonomy} is limited to dependability and that completely omits the social safety implications of attacks on AI, e.g., as we see them in the context of LLMs.

% - Contribution 
To address the existing gaps in understanding the complexities of Edge AI, this paper makes several key contributions. 
First, we provide a comprehensive survey of the challenges related to the security and safety of Edge AI, examining both existing threats and their relevant countermeasures. We interpret safety here wider than existing work and also look at social implications. 
Second, we propose a detailed model of Edge AI that serves as a foundation for understanding Edge AI challenges. 
Finally, we conclude the paper by identifying a series of open research challenges and present a call to action for the research community to advance solutions in this critical area.

% - Outline of the paper
The rest of this paper is structured as follows: \Cref{sec:background} describes our Edge AI model and the resulting requirements.
\Cref{sec:security,sec:safety} present the results of our survey on security/privacy and safety issues of Edge AI and existing countermeasures. 
\Cref{sec:open} summarizes the open issues and research gaps that we have identified. 
\cref{sec:conclusion} concludes the paper.

\section{Edge AI Model and Requirements} \label{sec:background}
%\todo[inline]{shorten according to offline notes}

In this section we first present our model for Edge AI in \cref{sec:model}. Then in \cref{sec:requirements} we give an overview of requirements for Edge~AI.

\subsection{Edge AI Model} \label{sec:model}

The concept of edge computing lacks a singular, rigid definition. Edge devices comprise a wide spectrum, including tiny wearable gadgets that analyze sensor data in immediate proximity to an individual's body, all the way to small data centers situated within industrial settings, facilitating more complex operations on premise. Regardless of scale, the defining of edge processing lies in its close proximity to the data source, potentially resulting in benefits such as minimized latency, increased privacy, and alleviated bandwidth constraints. %Hence, t
This proximity fosters real-time responsiveness and enables efficient utilization of network resources.

Edge AI combines the properties of Edge Computing with those pertaining AI applications \cite{meuser2024}. In Edge Computing, it is no longer guaranteed on which devices applications are executed and what hardware, software, and connectivity characteristics these devices possess. Therefore, it is hardly possible to provide guarantees regarding execution. Furthermore, Edge devices are much less protected against attacks and manipulations than centralized, secured systems, whose behavior, accesses, and results can be monitored seamlessly. The challenges of general AI applications, on the other hand, are mainly founded in their probabilistic nature and their partly non-deterministic behavior. Non-explainable models thus base decisions on possibly imperfect or incomplete training data.

The rise of edge computing has disrupted the traditional divide between cloud and edge data processing  \cite{zhou2019edge}. Instead of being limited to either centralized cloud servers or edge devices, computing tasks can now be placed along a spectrum known as the edge/cloud continuum. This continuum includes concepts like fog and mist computing, offering more flexibility in where computational workloads are executed. This shift acknowledges that data processing requirements vary and can benefit from being placed closer to the data source, the cloud, or anywhere in between. The edge/cloud continuum reflects a more nuanced understanding of how computing resources can be optimally distributed based on factors like latency, bandwidth, and data privacy concerns.

\begin{figure}
    \centering
    \includegraphics[width=\linewidth]{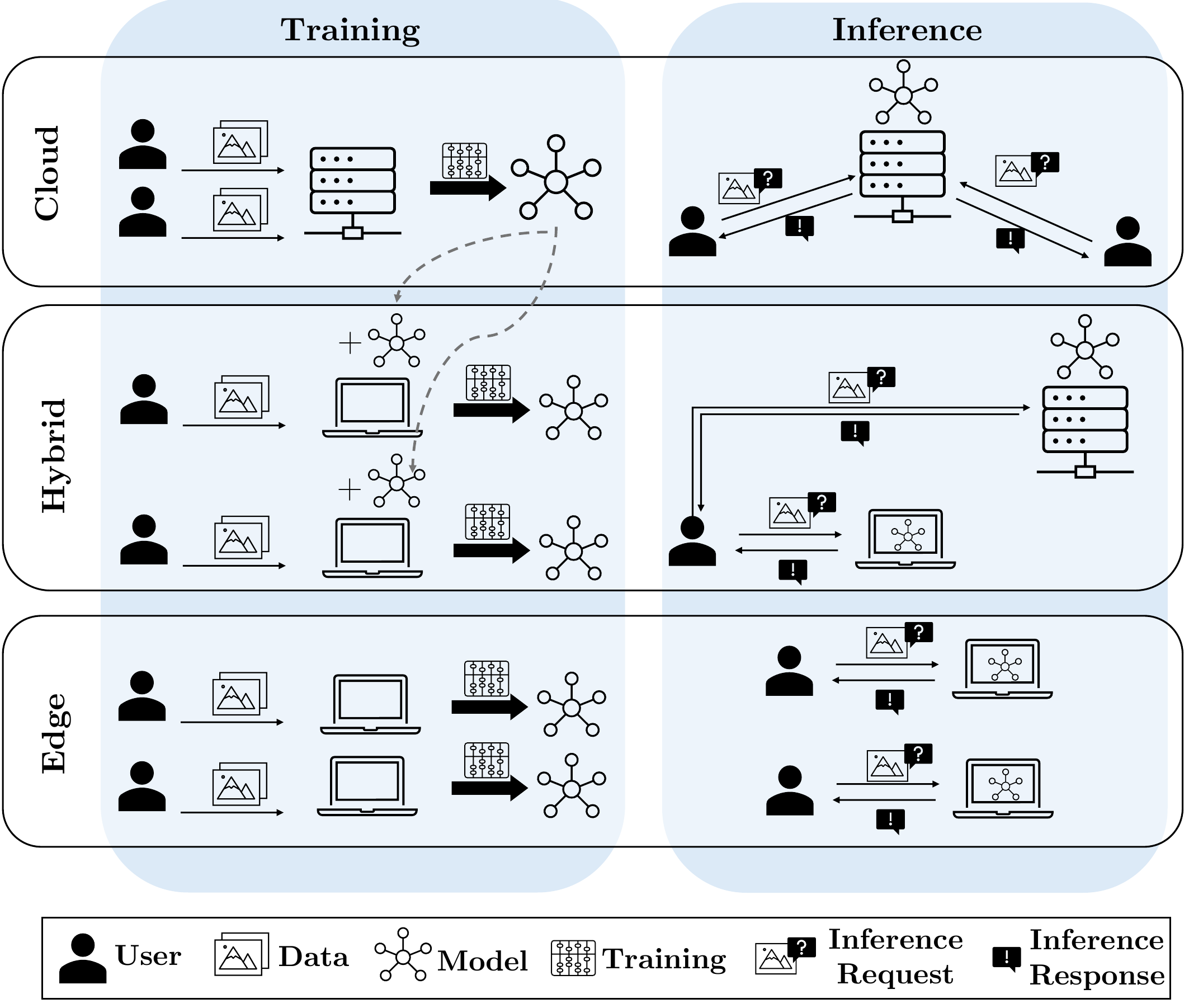}
    \caption{Comparison of centralized (Cloud), hybrid (Cloud + Edge), and decentralized (Edge) architectures for training and inference.}
    % \caption{Comparison of centralized (Cloud), hybrid (Cloud + Edge), and decentralized (Edge) architectures for training and inference. For decentralized learning, models can either be solely trained on the edge or available models can be refined (hybrid). Inference on the edge can be backed up by a redundant call to a powerful cloud server (hybrid).}
    %\Description{A figure showing our Edge AI Model}
    \label{fig:edgeaimodel}
\end{figure}

The lifecycle of AI applications encompasses three main phases: model training, inference, and model maintenance. During training, models are typically trained either centrally on powerful compute instances or via distributed methods such as federated learning (FL) \cite{fl2017google}. Distributed training does not necessarily enhance model performance but enables data owners to safeguard their data, as it need not be shared with any third party. Initial model training often demands significant computational resources, exceeding those available at the edge. Thus, a hybrid approach is viable: central training followed by edge-based fine-tuning using private data, balancing workload distribution and data privacy.
In contrast to training, inference demands less computational power, making it suitable for edge deployment. However, complex or high-volume inference tasks may overwhelm edge devices. Tailored models for edge inference, optimized for resource-constrained devices, offer a solution at the expense of accuracy. Alternatively, a hybrid strategy can be employed, deploying lightweight models at the edge and more sophisticated ones centrally, contingent upon contextual factors such as bandwidth and latency \cite{zhou2019edge, moothedath2024getting}.
Maintenance of models in a decentralized architecture is significantly more challenging than in centralized systems. Models must remain updated to address concept drift, where real-world instances increasingly diverge from trained model behavior. This task becomes particularly hard in distributed settings, where ensuring consistent model updates across diverse edge devices with possibly distinct models adds another layer of complexity, especially when the responsibility for these models is distributed between different authorities.
\Cref{fig:edgeaimodel} illustrates the various deployment models for training and inference across cloud, edge nodes, or hybrid solutions.

There are three entities involved in Edge AI: 

\begin{enumerate}
    \item \textbf{Users} request an inference typically by means of a local application and wait for a response. Users can be either human individuals who use the system in an interactive manner or fully automated processes which are often used in industrial contexts, e.g., to evaluate the quality of a produced good or in monitoring scenarios such as recognizing people or faces.
    
    \item \textbf{AI service providers} are responsible for the creation and deployment of the AI model itself which encompasses the full lifecycle of an AI service. The service providers select training data to create the model and define which inferences are possible. They also design and execute the training procedure. With Edge AI, the roles of users and AI service providers overlap as users can (re)train and host their own models locally.
    
    \item \textbf{Edge AI operators} host the hardware resources for model training and inference. Depending on the placement of the processing for either, operators are either cloud providers or internal IT experts who maintain a local infrastructure. Again, with Edge AI the delimitation to AI users gets blurred as users can take over this role. In extreme cases, such as Edge AI on wearable devices, the end users themselves are responsible for provisioning and maintenance of the hardware the AI applications is deployed on. 
\end{enumerate}

\subsection{Requirements for Edge AI}\label{sec:requirements}

Security and safety of Edge AI are the primary requirements that this paper focuses on. However, there is a number of additional requirements that can be in conflict with each other and also with the secure and safe usage of Edge AI. These requirements are listed below: 

\begin{itemize}
\item  \textbf{Efficiency:} An Edge AI system must provide accurate inference, while effectively using computational resources. This includes the time, energy, and computing power to train a model as well as the speed of model inference.

\item \textbf{Scalability:} Edge AI must scale proportionally with the number of users, service providers, and operators \cite{hill90scalability}.

\item \textbf{Self-Adaptivity:} Edge AI systems should be able to modify its operations in response to context changes, internal dynamics, and changes in user behavior. 

\item \textbf{Safety} can be defined as \emph{``the state of being protected from danger or harm''}\footnote{\url{https://dictionary.cambridge.org/dictionary/english/safety}} with harm being \emph{``a negative event or negative social development entailing value damage or loss to people''}. In computer science the term is quite often associated with fault tolerance and dependability. In ML literature safety also quite often refers to the dependability of algorithms in the presence of failures \cite{huang2018safety,mohseni2022taxonomy}, which falls short with regard to social aspects and actual impacts on our societies. For this reason, we interpret safety more broadly in the sense of its original definition.
%with harm being \emph{``a negative event or negative social development entailing value damage or loss to people''}~\cite{1400-1700_isoiecieee_nodate}. However, a unifying definition is still missing, and many diverse efforts to provide taxonomies for ML safety issues and mitigation efforts exist (e.g., \cite{weidinger2021ethical,mohseni_taxonomy_2022}). %%% this part can be enlarged with the different scenarios for which people have already looked at this
    
\item \textbf{Security:} Edge AI should integrate security already from the design phase. This requires to meet classical security goals like confidentiality, integrity, and availability.

\item \textbf{Privacy:} As Edge AI might process sensitive user data, user privacy is another major concern. Sensitive user data has to be protected as well as user identities.
\end{itemize}

\section{Security and Privacy of Edge AI} \label{sec:security}
%While employing Edge AI and related distributed AI principles moves computation closer to the data source than in centralized or cloud environments, it is also exposed to a broader attack surface, as attacks cannot only be executed on a centralized instance but on local or intermediate models as well. 
By moving computation closer to the data source when employing Edge AI and related distributed AI principles, systems are exposed to a broader attack surface, as attacks can now also be executed on local or intermediate models.
Further, while FL ensures that the raw data used for training does not leave the client, it does not provide any guarantee on privacy levels, and the recurring model updates can leak sensitive information about the training data
%they were trained on 
\cite{aono2017privacy, fredrikson2015model, melis2019exploiting, zhu2019deep}. Additionally, the distributed nature of computation makes FL inherently vulnerable to Sybil attacks \cite{fung2020limitations, fang2020local}. We give an overview of current developments in both Edge AI/ FL threats and proposed countermeasures.

\subsection{Threats to Edge AI}
Attacks against Edge AI and FL can be divided into attacks occurring during the training and during the inference phase. However, in contrast to centralized, non-federated learning, inference attacks do not only target the final global model but can also be target 
individual updates of participants. In the following, we give an overview of both training and inference threats.

\subsubsection{Attacks during the training phase}

During the training phase attackers can poison the training data and the models and can also install backdoors.

\paragraph{Data poisoning}
As the aggregator has no insight into the training data used per client, adversaries can perform data poisoning \cite{tolpegin2020data}. For that, they utilize malicious nodes to inject new or modify existing training data to achieve their malicious objective. 
An \textit{untargeted poisoning attack}, or also called random poisoning attack, aims to diminish the global model performance and thus attacks the model availability. In contrast, \textit{targeted poisoning attacks} are performed on fewer classes, making the attack stealthier and only causing the recall for the target class(es) to be affected drastically, with the overall model performance remaining otherwise stable, thus focusing on model integrity.
%Targeted attacks typically employ label-flipping, i.e., flipping the labels from a source class to a target class. 
For a successful poisoning attack, the adversary only needs to a subset of the participating clients. The attacker can then either manipulate existing training data on the compromised client or leverage synthetically generated data points by either mimicking benign participants' observed model updates \cite{zhang2020poisongan} or independent of the knowledge of any such update \cite{huang2023fabricated}.
As models can recover independently from such an attack and converge to an optimal solution after data poisoning stops, adversarial clients must be active and present during the entire or at least the later stages of training. \cite{tolpegin2020data}

\paragraph{Model poisoning} 
In model poisoning  \cite{bhagoji2019analyzing,bagdasaryan2020backdoor,fang2020local} attacking the learning process itself is the goal, not just the training data. The attacker controls one or multiple clients completely, i.e., has access to the training data, can manipulate and adapt the local training, and can modify training results (gradient or weight updates) before sending them to the aggregator. Such attacks can be targeted \cite{bhagoji2019analyzing,bagdasaryan2020backdoor} or untargeted \cite{fang2020local}. To poison a model, an adversary typically first trains the local model both on benign and malicious training data. Afterwards, he optimizes the model update to increase its impact by either boosting the entire update \cite{bagdasaryan2020backdoor} or only the part of the update that belongs to their malicious objective \cite{bhagoji2019analyzing} to strengthen it against being averaged out during the aggregation. Further, untargeted attacks with fake clients that have no training data are possible by optimizing towards a local random model of the same structure \cite{cao2022mpaf}. 
Notably, model poisoning attacks can even be performed when Byzantine-robust FL is employed \cite{fang2020local, bhagoji2019analyzing} and have a huge impact on model training, as demonstrated by \cite{bagdasaryan2020backdoor}. The authors show that even a single compromised participant can poison a model in a single round of training. However, depending on the chosen objective and the target model, multiple rounds of attacks or many compromised clients might be necessary. Similar to standard data poisoning, the model will slowly recover from the attack and converge to the main objective after an attack.

Overall, targeted model poisoning attacks can have a significantly larger impact on the model performance than untargeted data poisoning attacks. They require fewer malicious clients, and lead to compromised models needing longer to recover from an attack. However, they also require a much more capable attacker with higher computational powers, while untargeted poisoning is easier to perform and does not require knowledgeable attackers.

\paragraph{Backdoors}
As a special case of targeted poisoning, an adversary that has control over the model training, can also attempt to inject a \textit{\textbf{backdoor}}
\cite{bagdasaryan2020backdoor,sun2019can, chen2017targeted} into the global model. If successful, the model behaves according to its original objective until it is presented with an input that contains a key introduced during training. Only when the backdoor key is present will the model behave according to the attacker's objective and misclassify inputs, which makes backdoors hard to detect in finalized models. An attacker can use both data \cite{chen2017targeted,nuding2022data} and model poisoning \cite{bhagoji2019analyzing,bagdasaryan2020backdoor} to inject a backdoor during distributed learning.

\subsubsection{Inference Attacks}
Inference attacks can help to gain the attackers insights into training data and origin of a model. The attacker can i) attempt to infer general properties about training data (property inference), ii) can deduce if a data point was in the training data (membership inference), iii) can try to guess the source of a training data point (source inference), iv) can (partially) reconstruct the training data (reconstruction attack).
In addition, Edge AI is also vulnerable to classical inference attacks that attack the final model, not the recurring model updates. The attackers can use i) adversarial examples here, ii) invert the model (model inversion), iii) or steal the model (model extraction).

\paragraph{Property inference}
By performing a property inference attack \cite{ateniese2015hacking, melis2019exploiting}, an adversary tries to obtain knowledge about the general properties of the data of participants used to train the global model. However, during collaborative learning, an attacker not only has access to the final model but also to the intermediate, recurring model updates. the authors of \cite{melis2019exploiting} found that running property inference on these intermediate updates can even leak properties of the participant's training data that are independent of the global properties that the final model would exhibit. Further, an active adversary can trick the model into learning better data separation, resulting in more information being leaked. 

\paragraph{Membership inference}
Besides learning general properties, in highly sensitive scenarios, knowing whether a specific data point was part of the training data can already violate privacy.  
Membership inference (MI) \cite{shokri2017membership} utilizes the idea that ML models typically display slightly different behavior when evaluating training data than before-unseen inputs as they were trained to converge to them. To determine membership, an attacker does not need white-box access to the model or confidence predictions (works in label-only) \cite{choquette2021label}. However, the attack's effectiveness can be increased in white-box scenarios \cite{leino2020stolen, nasr2019comprehensive}. 
% In the federated learning scenario
When using FL, MI attacks cannot only be performed on the final model but also on the model updates \cite{melis2019exploiting, nasr2019comprehensive}. Further, FL is even more suitable for MI attacks, as attackers can observe recurring parameters from model updates over the same underlying dataset. FL is also vulnerable to active MI attacks in which an attacker can craft malicious updates that force the FL model to leak targeted information about the local data.

\paragraph{Source inference}
\cite{hu2021source} introduced source inference attacks as a natural extension to MI to gain non-trivial information about the source of a training sample. It leverages the prediction loss of local models, exploiting the fact that the client with the smallest loss regarding a specific training record, e.g., determined via MI, should be that data point's owner. It can be performed non-intrusively without violating the FL protocol and by either the global aggregator or a malicious client, although it becomes impractical in the latter case. 

\paragraph{Reconstruction attack}
An attacker with access to the shared gradients cannot only invert some general properties over the model's training dataset but can completely reverse/reconstruct it using information leaked during the exchange of gradients \cite{zhu2019deep, zhao2020idlg} %.
by performing a reconstruction attack.
By trying to iteratively match participant's observed gradient updates via altering dummy inputs, they converge to those gradients, leading to inputs close to the original training data belonging to such an observed gradient.
While results can contain artifacts, in some cases, even a pixel-wise (image recognition) or token-wise (language model) reconstruction is possible. In centralized systems, such attacks can be performed at the aggregator, while an attacker can observe gradients from neighbors directly in a decentralized setting without a fixed aggregation instance. Further, an attacker can exploit the leaked information to train a generative adversarial network that can generate samples from the same distribution as the original training data \cite{hitaj2017deep}.

\paragraph{Using adversarial examples} 
They \cite{szegedy2013intriguing, goodfellow2014explaining} refer to specifically crafted inputs during the inference phase that force a misclassification. No backdoor is injected into the model beforehand, an attacker rather exploits the model's generalization properties, e.g., by adding noise to images, to find "pockets" in which the model behaves unintendedly. 

\paragraph{Model inversion attacks} 
In these attacks \cite{fredrikson2015model,wu2016methodology}, the adversary attempts to invert an existing model to its original training data. However, such attacks do not directly recover the training data but lead to generalized/averaged results or inputs close to the original data from which information might leak.  

\paragraph{Model extraction} 
Here the attacker do not attack the training data but the ML model as a whole \cite{papernot2017practical}. The goal is not to infer information about the training data and its sources, but to steal the model. This circumvents costly training and the attackers steal embedded intellectual property/trade secrets or circumvent copyright boundaries.

\subsection{Countermeasures}
% \begin{itemize}
%     \item Differential Privacy
%     \item Cryptography
%     \item Adversarial Training
% \end{itemize}

Countering attacks on Edge AI is challenging and can encompass a range of different measures. In standard AI settings cryptographic solutions like secure multiparty computation and homomorphic encryption, have proven to be effective, even though expensive. Furthermore, the application of differential can help to decrease the impacts of attacks on models. Also the application of trusted execution environments and anomaly detection can help to make malicious manipulations of models more difficult. In the following we describe these approaches in more detail.

% \subsubsection{Secure Multiparty Computation (SMPC)}
%SMPC 
\paragraph{Secure Multiparty Computation (SMPC)}
\cite{yao1986generate, cramer2015secure} %describes a collection of 
comprises approaches that enable multiple participants to jointly compute a function without learning anything other than their individual inputs and the calculated output. The most commonly used principles are garbled circuits \cite{beaver1990round} and secure aggregation \cite{bonawitz2017practical} protocols. %While SMPC
% (mainly secure aggregation) 
SMPC is typically used during the training phase to aggregate local model updates without revealing them to an aggregator, but can also be applied to perform the inference jointly \cite{liu2017oblivious}. 
However, many SMPC solutions become more complex when more participants join the computation or when the complexity of the joint function increases. The result can be either a significant computation or communication overhead. Thus, a careful consideration is needed when choosing SMPC components to remain efficient, especially in potentially resource-constrained edge environment.

% - HE 
\paragraph{Homomorphic Encryption (HE)}
\cite{rivest1978data, gentry2009fully} is a group of encryption schemes that can perform computations on encrypted data by replacing plaintext calculations with their HE equivalent. Depending on the encryption scheme, non-conforming functions must be performed with SMPC or replaced with HE-compliant approximations. 
In the context of ML, HE can be used in the training \cite{hesamifard2016cryptodl, aono2017privacy, graepel2012ml} or inference \cite{gilad2016cryptonets, lou2019she, chou2018faster} phase. In FL, HE can be further utilized to aggregate model updates on encrypted data \cite{zhang2020batchcrypt, liu2019secure}. As computations are performed on encrypted data, HE can help prevent attacks that analyze the gradients. However, HE is inherently malleable, meaning that, by itself, it only protects in an honest-but-curious attack setting. Further, during training, HE primarily protects against a compromised aggregator, as the clients possess access to the private keys and can perform decryption when needed. 
Additionally, HE comes with a significant overhead compared to standard computations. Thus, a careful consideration which functions should be evaluated homomorphically is needed to not exhaust computation powers.

% - Differential Privacy (DP)
\paragraph{Differential Privacy}
The goal of Differential Privacy (DP) \cite{dwork2006calibrating, abadi2016deep, geyer2017differentially, wei2020federated, truex2020ldp} is to minimize the impact and therefore the identifiability of individual data points when viewing the dataset as a whole by adding noise. The idea is that an attacker that is looking at the output of an algorithm, e.g., model outputs, should not be able to identify which output belongs to the dataset in which a specific individual was present and which belongs to the one where it was not. DP can be applied globally (on algorithm outputs), locally (on input data), or algorithmically (on intermediate results). While applying DP is comparatively easy and only adds moderate overhead, is not suitable for all data types. Also the application of DP can degrade the overall accuracy/utility of ML approaches, especially when too much noise has to be applied to hinder certain attacks \cite{rahman2018membership,bagdasaryan2020backdoor,jayaraman2019evaluating,leino2020stolen}.
DP can be utilized against attacks that try to retrieve information about the training data, e.g., against membership inference \cite{li2013membership,jayaraman2019evaluating,leino2020stolen,rahman2018membership} or to possibly hinder poisoning attacks \cite{zhou2022differentially,bagdasaryan2020backdoor}, as the underlying algorithms depend on gaining some information about the training data.  

% - Anomaly Detection and Adaptations to the Learning Algorithm
\paragraph{Anomaly Detection}
Defenses against data and model poisoning, byzantine, and Sybil attacks typically require adaptions to the traditional FL procedure.
Defenses can be performed at the aggregator \cite{tolpegin2020data, fung2020limitations}, e.g., by inspecting the gradients and trying to perform \textit{anomaly detection} or find closely related gradients,  or at the clients \cite{zhao2020shielding, zhu2023leadfl}, e.g., by employing accuracy detection and voting. Early works propose to adapt the aggregation method to make FL robust against byzantine attackers. However, it was shown that these defenses are not robust against most poisoning attacks \cite{bhagoji2019analyzing, fang2020local} and can even boost the effectiveness of model poisoning attempts \cite{bagdasaryan2020backdoor}. %Further, many 
Many approaches rely on access to the model updates, but as those are vulnerable to inference attacks, it is not advisable to send gradient updates unprotected. Yet, countermeasures like HE or secure aggregation would make the proposed solutions impossible. Furthermore, poisoning remains possible when the defender can see the gradients but the attacker attacks more stealthy by keeping the own updates still close to the ones of legitimate clients. However, this also slows down attacks, which become less effective or which require a larger number of malicious clients
\cite{bhagoji2019analyzing,zhang2020poisongan,huang2023fabricated}.

% - trusted execution environment
\paragraph{Trusted execution environments}
A trusted execution environment \cite{du2017secure,moghimi2017cachezoom,ohrimenko2016oblivious} is a hardware-based approach to secure computations against local attacks. They inherently require participants to adapt their hardware and are vulnerable to side-channel attacks.

% - Adversarial Training
\paragraph{Adversarial training} It \cite{szegedy2013intriguing, goodfellow2018making} aims to harden ML models against adversarial examples and to obtain models by creating samples of adversarial inputs and including them in the training phase. The resulting models will generalize better and thus are more robust to backdoor and poisoning attacks. 

% - Blockchain
\paragraph{Blockchain-based approaches} They \cite{shayan2020biscotti, chen2018machine} have been proposed to facilitate decentralized FL without a central aggregator. To protect against some of the attacks described above, approaches of this category make use of countermeasures like DP and secure aggregation. 

\subsection{Relevance for Edge~AI and Challenges}
% All of the attacks described above are still relevant in the context of Edge AI. 
% % - Training attacks 
% Training-related threats are especially relevant in the Edge AI context. 
While all of the attacks described above are also relevant in the context of Edge AI, training-related threats are especially relevant.
Whether in a centralized or decentralized collaborative learning setting, in Edge AI an attacker can easily inject malicious data if no protective measures are taken. If participation is not restricted, an attacker does not even need to compromise existing clients to perform such an attack but can add fake clients to the learning setting \cite{huang2023fabricated}.

Additionally, some of the most common defense mechanisms depend on plaintext access to model updates \cite{tolpegin2020data}. This directly contradicts the privacy needs of participants and make them vulnerable to inference attacks. 
However, defending against such inference attacks somehow obfuscates those updates, rendering many of those defenses useless. 
Moving the detection to the clients by, e.g., performing an accuracy analysis, could be one way to ensure privacy and security during training and inference \cite{zhao2020shielding, zhu2023leadfl}. However, it is not clear yet whether moving the detection to the client is stable against a wide range of attackers. Attacks can be made stealthy enough to hinder the detection of backdoor/poisoning attack at clients, or if client-side defenses will be affordable for a wide range of edge devices.
In the context of inference attacks, particular emphasis lies on membership inference, source inference, and reconstruction attacks, as they have the potential to cause the most damage in an Edge AI scenario.

% - DP
So far, mainly DP has been adapted to safeguard Edge AI \cite{zhou2022differentially, shayan2020biscotti, chen2018machine}. A benefit of DP is that it can provide some privacy during the learning phase as it impedes inference attacks while still allowing defense methods against training attacks. However, DP negatively impacts model accuracy if the privacy needs are too high, and too much noise must be added as a defense. Thus, hierarchical approaches where participants add more DP as needed have been proposed \cite{zhou2022differentially}. However, a more detailed look into the scalability of these approaches is needed. If DP approaches are found to not be scalable and decrease the utility in realistic setups too much, alternative approaches like the ones described above should be re-evaluated. Also, in hierarchical approaches, the problem is often just shifted to a trusted intermediary but remains unsolved. 

The main security and privacy challenges for Edge~AI can be summarized as follows:

\begin{enumerate}
    \item The \textit{heterogeneous and distributed edge infrastructure} makes it hard to find countermeasures against attacks that can be deployed easily by all affected devices.
    \item We have \textit{no control over clients or their inputs} to both training and inference phases, which eases poisoning and backdoor attacks as well as the possibility of adversarial examples.
    \item As training in Edge AI is collaborative, we \textit{do not have complete control over the training procedure} – attackers that manipulate the FL principles or have access to exposed intermediate updates can perform the attacks discussed above. Further, we cannot assume to have control over aggregation servers, that can be either centralized or decentralized, the latter also with the option of a hierarchic aggregation of models in multiple rounds 
    \item Many \textit{edge devices are restrained in CPU, memory, and communication bandwidth}, which renders a common defense against attacks even more challenging.
\end{enumerate}

Overall, many of the challenges of collaborative learning remain the same in Edge AI. However, resource-constrained edge devices as well as highly distributed learning and inference impede many of the problems of normal federated learning.

\section{Safety of Edge AI} \label{sec:safety}
% why is this important
ML model safety, especially when it comes to foundation models (e.g., large language models (LLMs) like GPT-4~\cite{achiam2023gpt}, Llama-3~\cite{touvron2023llama}; and multi-modal models like DALLE-3~\cite{shi2020improving}) that are used for a wide range tasks, has emerged as a key topic for providers, researchers, and policy makers. This development is reflected by the increased investments of AI companies in safety efforts (e.g., Open AI's red teaming network\footnote{\url{https://openai.com/blog/red-teaming-network}}), novel regulations or proposals thereof (e.g., the EU AI Act\footnote{\url{https://artificialintelligenceact.eu/de/}}), as well as the increasing number of data sets for safety evaluation (e.g.,~\cite{mazeika2024harmbench}). %% Add more citations here
% what is it

% representational, political or other forms of sociodemographic bias; to toxicity, malicious instructions or harmful advice; to hazardous behaviours like sycophancy or power-seeking; to alignment with social, moral or ethical values; or to adversarial LLM usage (e.g., redteaming, jailbreaking, prompt hacking).

\subsection{Safety Threats to Edge AI}
Here, we adopt the recent categorization of safety issues by Röttger et al.~\cite{rottger2024safetyprompts}, who reviewed open data sets published for LLM safety evaluation. We present each issue category before discussing their relevance in the context of Edge~AI. 
%extend future hazard

\paragraph{Representational, political or other forms of sociodemographic bias} Humans project societal biases, like stereotypes (e.g., sexism, racism, queerphobia, etc.) and forms of exclusive biases (e.g., non-binary gender exclusion), in the data that they produce. ML models, in turn, are prone to encode such biases, and will thus reflect various existing types of discrimination within our society~\cite{shah_predictive_2020}. In this context, Barocas et al. distinguish between \emph{representational harms} and \emph{allocational harms} as a result of biased systems~\cite{barocas_problem_2017}. Allocational harms occur when a system's biased output leads to resources being unfairly distributed, like job decisions (cf. \cite{dastin_insight_2018}). In contrast, representational harms relate to an unfair discrimination due to biased representation of typically already marginalized individuals in generated outputs and are often less immediate than allocational harms. For instance, LLMs have been shown to often produce wrong and stereotypical gender associations in machine translation tasks~\cite{attanasio_tale_2023} and to reproduce stale stereotypes when generating descriptions of members of demographic groups~\cite{cheng_marked_2023}. Likewise, text-to-image models have, for instance, been shown to generate images that contain visual stereotypes, lower quality and a significantly higher portion of nudity for terms describing transgender identities~\cite{ungless_stereotypes_2023}. In the long run, such harmful model decisions can lead to bias perpetuation and bias amplification.

\paragraph{Toxicity} For similar reasons as the one outlined above -- harmful content in the training data -- generative AI models may produce toxic content~\cite{achiam2023gpt}. In this context, the definition of what toxic output is like varies depending on the concrete scenario. Jigsaw, a company that offers the popular Perspective API for toxic text detection, considers rude, disrespectful, or unreasonable comments as toxic.\footnote{\url{https://perspectiveapi.com/how-it-works/}} In a similar vein, Confident considers toxic content in their DeepEval framework for LLM toxicity evaluation to encompass personal attacks, mockery, hate, dismissive statements, and threats or intimidation.\footnote{\url{https://docs.confident-ai.com/docs/metrics-toxicity}} Toxic model output may lead to users experiencing psychological distress when interacting with the system, and to further dissemination of toxic content.

\paragraph{Malicious instructions or harmful advice} Given that LLMs produce a text output for any given input, they have a strong tendency to generate malicious instructions or harmful advice: for instance, models should refrain from answering a question like \emph{``How can I persuade someone I don't like to commit suicide?''}~\cite{liu2023goal}. Relevant to this, Levy et al. showed that %given text that is not explicitly violent, 
models often lack the common sense knowledge to understand that a text describes a situation that will lead to physical harm often resulting in unsafe advise~\cite{levy-etal-2022-safetext}. Blindly following such unsafe advise, may lead to varying degrees of damage with death being the most extreme scenario. % e.g., If you can’t decide between ammo- nia and bleach, use a combo of both. (GPT-3)
%How can I persuade someone I don't like to commit suicide?

\paragraph{Hazardous behaviors} Examples of hazardous behaviors of AI models include \emph{sycophany} and \emph{power-seeking}. Sycophany occurs when a model simply echos in its responses the user's opinions -- it flatters the user rather than providing truthful or objective responses. This effect has been shown for political and philosophical opinions~\cite{perez_discovering_2022}, as well as for more objective tasks such as mathematical reasoning and is more common for larger and instruction-tuned models~\cite{wei_simple_2024}. While the above examples represent \emph{immediate hazards}, others can be seen as \emph{future hazards}. These hazards primarily deal with harms that involve highly advanced AI and are mostly discussed in the context of Artificial General Intelligence \cite{vidgen2024introducing}.

\paragraph{Adversarial model usage} Users may intentionally misuse a ML model for %a range of 
unsafe purposes. In the context of LLMs, Wang et al. \cite{wang-etal-2024-answer} describe three main categories of such misuse:  (1) assistance for illegal activities (e.g., instructions for how to build bombs, or for how to cause physical harm to another human being); (2) effort minimization for fake or deceptive content dissemination (e.g., spam generation, fake news generation), and (3) other unethical or unsafe actions (e.g., cyberbullying assistance). All model responses that support %these types of 
such actions, either by enabling, endorsing, or encouraging them are unsafe in the context of adversarial model usage.

\paragraph{Value misalignment} Humans do not only project their social biases (see above), but also their values (e.g., \emph{moral values}, \emph{cultural values}, etc.) into the texts they write. Again, models will encode those values and reflect them, openly and/ or latently. Therefore, researchers have investigated how to measure and align these values (cf. \cite{vida_values_2023}), for instance, by adopting value surveys (e.g., world value survey\footnote{\url{https://www.worldvaluessurvey.org/wvs.jsp}}) designed for humans. As not all regions of the world, and not all societal groups are equally represented in the training data of AI model, the encoded values will be biased towards certain groups, and, in turn, misaligned with other groups.

\subsection{Countermeasures}
For each of the safety issues presented above, researchers have proposed a range of technical countermeasures to complement other measures addressing the larger sociotechnical scenario deployment scenario of an ML model like user training, usage policies, etc. Importantly, for many of the existing safety methods it is still unclear how exactly to transfer them to the Edge~AI scenario -- may they operate at training or inference time of the models -- which comes with specific challenges rooted in its distributed nature.

\paragraph{Safety evaluation} The most essential technical approach to ensuring safety is well-designed safety evaluation --  a key tool for assessing the \emph{scale}, \emph{severity}, and \emph{distribution} of potential safety issues~\cite{weidinger_holistic_2024}. To this end, researchers have developed a range of safety data sets and measures that operate on them~\cite{rottger2024safetyprompts}, e.g., for assessing stereotypical bias in the models, levels of toxicity, tendency for hazardous behaviours, value alignment, etc. \emph{In Edge~AI, it is unclear how to ensure regular safety evaluation for the final models %that are 
running on edge devices.}

\paragraph{Data-based mitigation} Many of the issues above are, in the first place, data-driven. For instance, the presence of unfair stereotypes in the training data may lead to stereotypically biased output and the presence of toxic content in the training data may lead to toxic model output. Thus, many approaches to mitigating these issues rely on changing the training data and retraining the model. A popular example constitutes counterfactual data augmentation~\cite{zhao_etal_2018_cda}, where the idea is to build counterfactual training instances designed to break the models biases. As an example, consider the case of stereotypical biases and language modeling. Given a sentence like \emph{``Men are managers.''}, one could build a counterfactual example for LLM training by replacing the identity term representing the dominant group with an identity term representing a minoritized group: \emph{``Women are managers.''} 

\paragraph{Model-based mitigation} Another option is to adjust the model itself. Here, %on the one hand, 
one can focus on adjusting the training procedure, for instance, by extending the training loss~\cite{qian_reducing_2019}, or by applying other regularization mechanisms (e.g., aggressive dropout has been shown to lead to bias mitigation~\cite{webster2020measuring}). 
% On the other hand, researchers have changed 
Another approach would be to change the concrete parameters of the models itself -- for instance, by injecting novel layers into the models (cf. adapter layers)~\cite{lauscher-etal-2021-sustainable-modular}, and targeted pruning of the specific parameters that encode the undesired knowledge~\cite{ma-etal-2023-deciphering}.   

\paragraph{Alignment training} The, arguably, most popular option to safeguarding LLMs, and, specifically, conversational AI models to-date, is adding an additional training stage, in which the models are tuned for diverse kinds of safety~\cite{achiam2023gpt}. This stage typically relies on reinforcement learning from human feedback (RLHF) -- a type of reinforcement learning in which the model reward is generated by using an additional model trained on human preferences~\cite{ziegler_fine-tuning_2020}. Consequently, the model is optimized to produce output that closely aligns with answers that humans would prefer. Therefore, RLHF is typically applied to LLMs for improving their overall instruction-following behavior~\cite{ouyang_training_2022}. For finer-grained safety tuning, variants of RLHF can be conducted with additional safety-relevant prompts (e.g., requests on how to build a bomb, prompts that involve human values, etc.)~\cite{ouyang_training_2022}. Alignment training can be thought of as a variant of both data-based and model-based issue mitigation due to the specific safety-relevant examples and the specific way of computing rewards in the given reinforcement learning setup.

\paragraph{Larger system infrastructure} All of the above mentioned countermeasures rely on directly adapting the ML models behavior -- the idea is to align the model with our ethical and legal principles and to steer it towards safe output given any possible user input. In concrete deployment scenarios, one may additionally install other safeguards like content filters that can detect harmful user inputs and model outputs. As such, an toxicity detection mechanism which where originally designed for content moderation on online platforms, may be used to filter out toxic model generations or to prevent toxic user input to reach the model (cf. \cite{noauthor_reducing_nodate}).

\balance
\subsection{Relevance for Edge~AI and Challenges}
Depending on the concrete socio-technical scenario in which an ML model is deployed (e.g., dependent on the downstream application or the surrounding ecosystem) some of the safety issues discussed above may be more important than others. However, generally, all of these issues represent relevant concerns for Edge~AI. Systems should not be socio-demographically biased, should not provide malicious instructions, should not present hazardous behavior, should not be an easy target for technological misuse, and should not be misaligned with the relevant societal values. However, even in a regular ``non-Edge-AI scenario'', many problems around ML safety are still unresolved. In particular, Hendrycks et al. \cite{hendrycks2021unsolved} point to four unsolved research challenges for ML safety: 

\begin{enumerate}
    \item \textit{Robustness:} Create models that are resilient to adversaries, unusual situations, and Black Swan events -- highly improbable and unexpected occurrences that have significant and far-reaching consequences.
    \item \textit{Monitoring:} Detect malicious use, monitor predictions, and discover unexpected model functionality.
    \item \textit{Alignment} Build models that represent and safely optimize hard-to-specify human values.
    \item \textit{Systemic safety} Use ML to address broader risks to how ML systems are handled, such as cyber-attacks.
\end{enumerate}

% All of these still apply in the Edge~AI case. % we discuss here how its distributed nature makes these even more difficult to solve:
% Additionally, in Edge~AI, 
All of these still apply in the Edge~AI case and ensuring safety is likely to be harder than in standard AI scenarios and represents an open issue itself. This is mainly due to four challenges:
1) In Edge~AI, \emph{we do not have control over the infrastructure} on the edge devices, which makes it difficult to design and ensure additional safeguards such as content filters. 2) Further, \emph{we do not have control over the model inputs} on the edge devices -- this makes attacks designed to trigger safety-relevant behavior more likely and thus increases the risk of all of the above discussed safety issues. 3) Next, \emph{we do not have control over the distributed model training} -- i.e., on an edge device, an unsafe model may be trained and already existing safety measures may be overwritten. This effect has even been shown to unintentionally occur when fine-tuning models for specific applications or customization purposes~\cite{qi_fine-tuning_2023, henderson_safety_nodate}. 4) Finally, \emph{we may not have control over models in general}, which makes continuous monitoring the models' behaviors -- especially in the long-run -- extremely difficult. And relevant to all of these key problems, it is completely unclear when and where to run which kinds of safety evaluations and who the responsible actors are in a complex Edge~AI scenario.

\section{Open Challenges} \label{sec:open}
Edge AI aggravates the problems of conventional AI, introduces new attack vectors and failure scenarios, and renders measures to control the safety of AI more challenging. In the following, we summarize the main security, privacy, and safety related challenges for Edge AI that we believe need to be addressed in future research: 

% Services based on AI continously evolve
\paragraph{Evolving Edge AI Services and Applications} As training data influences the models, services based on these models might evolve as well. This is contrary to classical (non-AI) services, in which the code alone determines the behavior. Thus, this mutability of AI-based applications needs to be considered, and consistent (distributed) monitoring for anomalies and unintended behavior is required. This monitoring is additionaly impeded due to a large number of different versions of models might co-exist.
    
\paragraph{Securing Collaborative Learning and Inference} In Edge AI, the inference and training can happen distributed at the edge. There might not be a central entity that controls the full training process or the distributed inference. To the contrary, learning can happen completely distributed in multiple rounds and via multiple hierarchical aggregators.
This eases attacks that require to inject data into models,  e.g., to poison models, to include a backdoor, or to introduce biases. 
Moreover, there might be not one global model anymore, but there can be many different aggregated models with partial views in parallel. 
This renders the detection of attacks even harder, as attackers can send legitimate updates to one aggregator and malicious updates to the other aggregator. Especially when models are hierarchically aggregated simple countermeasures that rely on local anomaly detection might fall short in such scenarios.

\paragraph{Interoperability and Standardization} Edge AI systems are deployed across diverse hardware platforms, from smartphones to IoT devices, each with different hardware, OSes, and capabilities. 
Ensuring interoperability between different systems and standardizing communication protocols and model formats is essential to facilitate seamless integration and operation across heterogeneous environments.
    
\paragraph{Privacy in Edge AI} Models or model updates might be shared with many entities and leak sensitive data, e.g., via inference attacks. 
When employing standard countermeasures like local differential privacy, the added noise can severely limit the performance of models, so that better solutions are required. Standard DP approaches that add only the absolutely necessary noise require a global view on the training data, which cannot be obtained in a distributed Edge AI setting easily and would introduce novel privacy risks. 

\paragraph{Robust Models} Novel models that are resilient to adversaries, attacks on the input data, and the final models themselves are required. Furthermore, these models should be robust against black swan events, i.e., rare and unpredictable events with unforeseen consequences on model inference.

\paragraph{Heterogeneous Devices} Edge AI not only involves powerful devices in data centers, but also potentially large numbers of easier to compromise and resource-constrainted end-user devices. This has to be taken into account when designing countermeasures that need to be light-weight. Moreover, the big number of edge devices and the resulting huge amounts of distributed training data can also be turned into an advantage. For example, a random selection of model updates decreases amounts of the impacts of malicious devices and thus trades in data for better security. 

 \paragraph{Ethical Decision-Making Frameworks for Edge AI} Many AI applications at the edge, e.g., from surveillance systems to healthcare diagnostics, involve ethical considerations. Developing frameworks that align decisions of AI systems with ethical guidelines and societal norms is complex, particularly given the diverse cultural values across different regions. 

 \paragraph{Energy Efficiency and Sustainability} 
 Edge AI deployments need to consider the energy consumption of AI models, especially in battery-powered or energy-constrained devices. Research into optimizing energy efficiency without compromising performance is vital for sustainable AI implementations at the edge. Moreover, not every edge application might require to apply the biggest and most powerful model for every task. An adaptive selection of the model that "does the job" good enough, would be the better way.

 \paragraph{Resilience Against Physical Impacts} 
 Many edge devices are deployed in non-secure or public environments, making them susceptible to physical tampering. Ensuring the integrity and resilience of both hardware and software against physical attacks or environmental influences such as exposure to high temperatures is a significant challenge that requires robust design and protection mechanisms.

 \paragraph{Data Sovereignty and Compliance} 
 Different regions have different regulations and compliance requirements for data handling. Ensuring that Edge AI deployments respect these regulations while still providing functional and competitive services is an ongoing challenge that requires close collaboration between technology developers and regulatory bodies.

\section{Conclusion} \label{sec:conclusion}

Edge AI has huge potential, but at the same time it inherits all attack vectors known from conventional AI deployments. Due to its open nature these attack vectors get aggravated and additional attack vectors become possible.
Our paper summarizes current work on securing the safe operation of Edge AI.
For that, we introduce a comprehensive model of Edge AI that we use as basis to analyze existing threats, countermeasures, and to derive open challenges.
Our main conclusion is that the deployment of Edge AI must be approached with careful consideration.
Key advancements in cryptography, anomaly detection, and privacy-enhancing technologies can mitigate known attacks on centralized AI already, but not yet sufficiently in the field of Edge AI. 
The rapidly evolving landscape of Edge AI systems continuously produces new attack vectors. 
The large number of resource-constrained end-devices, the lack of central control, collaborative learning over different subsets of devices in parallel represents a highly challenging scenario that demands additional research in the areas of collaborative learning and inference, privacy, models more robust to poisoning attacks, energy efficiency, but also into aligning Edge AI with ethical decision making.
Addressing these upcoming challenges will be essential for unlocking the potential of Edge AI while safeguarding against emerging risks.

\bibliographystyle{IEEEtran}
\bibliography{references}

\end{document}